\documentclass[pra,twocolumn,showpacs,preprintnumbers,amsmath,amssymb]{revtex4}

\usepackage{mathrsfs}
\usepackage{txfonts}
\usepackage{graphicx}
\usepackage{dcolumn}
\usepackage{bm}
\usepackage{amsmath}

\begin{document}


\title{Controlled generation of field squeezing with cold atomic clouds coupled to a superconducting transmission line resonator}

\author{Peng-Bo Li}
\email{lipengbo@mail.xjtu.edu.cn}
\author{Fu-Li Li}
\affiliation{MOE Key Laboratory for Nonequilibrium Synthesis and Modulation of Condensed Matter,\\
Department of Applied Physics, Xi'an Jiaotong University, Xi'an
710049, China}


\begin{abstract}
We propose an efficient method for controlled generation of field squeezing
with cold atomic clouds trapped close to a superconducting transmission line resonator. It is shown that, based on the coherent strong magnetic coupling between the collective atomic spins and microwave fields in the transmission line resonator, two-mode or single mode field squeezed states can be generated through coherent control on the dynamics of the system. The degree of squeezing and preparing time can be
directly controlled through tuning the external classical fields. This protocol may offer a promising platform for implementing scalable on-chip quantum information processing with continuous variables.
\end{abstract}

\pacs{42.50.Dv, 03.67.Bg, 42.50.Pq} \maketitle

Squeezed fields have both fundamental and practical implications in
quantum optics \cite{1} and quantum information processing \cite{2,3,4,5}. Controlled generation of field squeezing in the microwave or optical range has been investigated in various physical systems such as cavity QED \cite{6,7,8,9,10,11} and superconducting quantum circuits \cite{12,13,14}. Solid-state superconducting quantum circuits are
believed to possess the advantages of integration and scaling in a
chip \cite{15}. Ultracold atoms are very attractive in view of long coherence times and the well-developed techniques for detecting and manipulating the ground electronic
(hyperfine) states. Hybrid systems consisting of ensembles of atomic or molecular system and
superconducting transmission line resonators have been intensively investigated \cite{16,17,18,19,20,21,22,23,24,25}. Recently, J. Verdu and coauthors have
shown that strong magnetic coupling of an ultracold atomic gas to a superconducting waveguide cavity is possible \cite{19}. This interesting work opens up the possibility for utilizing this hybrid system in the field of quantum optics and quantum information.
The present work is to design a scheme for producing squeezed state of the transmission line cavity fields using the hybrid system, which may be used
to implement on-chip quantum information processing with continuous variables.

In this brief report, we propose to controllably generate two-mode
or single mode squeezed states of the electromagnetic fields confined in a superconducting transmission line resonator. It
is shown that, based on the strong magnetic coupling of a cold atomic gas to a transmission line resonator \cite{19,20,21}, under certain conditions the coupled system of atomic spins and cavity modes can behave as three coupled harmonic oscillators with controllable coefficients. Through coherent control on
the dynamics of the system, at some instants the atomic spins are decoupled from the cavity modes, leaving the cavity fields in a squeezed state. The distinct advantage
of this method comparing to the traditional intra-cavity parametric amplification is that the degree of squeezing and
preparation time can be efficiently controlled through tuning
the intensities and detunings of external classical fields. Comparing to other proposals based on electric-dipole coupling between atoms and fields in cavity QED \cite{6,7,8,9,10,11}, this
scheme, utilizing strong magnetic coupling of atomic spins to cavity modes \cite{19}, has the advantage of being immune
to charge noise and could constitute qubits with much
longer coherence times. Combined with the technology of atomchip and circuit QED, our study may open promising perspectives for
the implementation of on-chip quantum information processing with continuous variables
in the microwave regime.

This proposal consists of an ensemble of N cold atoms trapped above the surface of a
superconducting transmission line resonator \cite{16,17,18,19,20}, as
sketched in Fig. 1. 
\begin{figure}[h]
\centerline{\includegraphics[bb=107 527 512 659,totalheight=1.1in,clip]{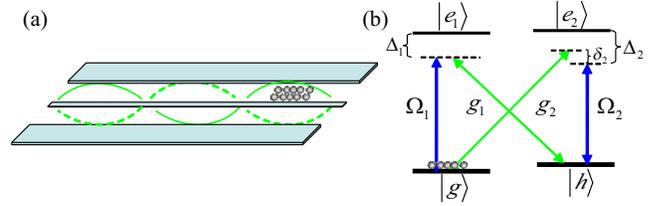}}
\caption{(Color online) (a) Ensembles of ultracold atoms trapped near the surface of
a superconducting transmission line resonator. (b) Atomic level structure with couplings to the cavity modes and driving laser fields.
The atoms are initially cooled in the ground state $\vert g\rangle$.}
\end{figure}
The transmission line resonator consists of three conducting stripes: the
central conductor plus two ground planes. The electromagnetic field of the resonator is confined
near the gaps between the conductor and the ground planes. A cloud of cold atoms can be trapped close to
the resonator by a variety of macroscopic electrostatic traps or other atomchip technology.
Two classical fields of frequencies $\omega_1$
and $\omega_2$ drive dispersively the atoms, establishing a couple
of Raman system through two cavity modes of frequencies of
$\nu_1$ and $\nu_2$. The two ground states of the atoms are labeled as
$\vert g\rangle$, and $\vert h\rangle$, and the
intermediate states as $\vert e_1\rangle$ and $\vert e_2\rangle$. The intermediate states $\vert e_1\rangle$ and $\vert e_2\rangle$
can be replaced by a single level \cite{6}, provided the two Raman channels remain distinct. The
classical fields drive dispersively the transitions $\vert
g\rangle\leftrightarrow \vert e_1\rangle$ and $\vert
h\rangle\leftrightarrow \vert e_2\rangle$ with Rabi frequencies
$\Omega_1$ and $\Omega_2$. The cavity modes couple the transitions
$\vert h\rangle\leftrightarrow \vert e_1\rangle$ and $\vert
g\rangle\leftrightarrow \vert e_2\rangle$ with coupling constants
$g_1$ and $g_2$. The detunings for these transitions are
$\Delta_1=\omega_{e_1g}-\omega_1=\omega_{e_1h}-\nu_1$,
$\Delta_2=\omega_{e_2h}-\omega_2$, and
$\delta_2=\Delta_2-\omega_{e_2g}+\nu_2$. Then in the interaction
picture and under the dipole
and rotating wave approximation, the Hamiltonian  describing this case is given by (let
$\hbar=1$)
\begin{eqnarray}
\label{H1}
H_I&=&\Omega_1\sum_{j=1}^N\vert e_1^j\rangle\langle g^j\vert e^{i\Delta_1t}+\Omega_2\sum_{j=1}^N\vert e_2^j\rangle\langle h^j\vert e^{i\Delta_2t}\nonumber\\
&&+g_1\hat{a}_1\sum_{j=1}^N\vert e_1^j\rangle\langle h^j\vert e^{i\Delta_1t}+
g_2\hat{a}_2\sum_{j=1}^N\vert e_2^j\rangle\langle g^j\vert e^{i(\Delta_2-\delta_2)t}+\mbox{H.c.},\nonumber\\
\end{eqnarray}
where $\hat{a}_i$ is the annihilation operator for the cavity mode with frequency $\nu_i (i=1,2)$.
We consider dispersive
detunings
$|\Delta_1|,|\Delta_2|,|\Delta_1-\Delta_2|\gg|\Omega_1|,|g_1|,|\Omega_2|,|g_2|$.
Since levels $\vert e_1\rangle$ and $\vert e_2\rangle$ are coupled
dispersively with both levels $\vert g\rangle$ and $\vert h\rangle$,
they can be adiabatically eliminated. Then the interaction Hamiltonian describing the coupling of the cold atoms to the cavity modes is \cite{6}
\begin{equation}
H_{I}=(\beta_2\hat{a}_2+\beta_1\hat{a}_1^\dag)\hat{c}^\dag+\mbox{H.c}.,
\end{equation}
where $\beta_i=\frac{\sqrt{N}\Omega^*_ig_i}{\Delta_i}$, $\hat{c}=1/\sqrt{N}\sum_{j=1}^N\vert g_j\rangle\langle h_j\vert$.
For weak
excitations of the atoms, the operator $\hat{c}$ obeys
approximate harmonic oscillator commutation relations
$[\hat{c},\hat{c}^\dag]\simeq1$ \cite{22,24,25}. In such a case, we can describe the
atomic excitations as a set of harmonic oscillator states
$\vert\textbf{0}\rangle_a=\vert g_1g_2...g_N\rangle_a$,
$\vert\textbf{1}\rangle_a=\hat{c}^\dag\vert\textbf{0}\rangle_a$,
etc. We assume $\beta_1=i\xi_1,\beta_2=i\xi_2,|\xi_2|>|\xi_1|$, and introduce $\Theta=\sqrt{|\xi_2|^2-|\xi_1|^2}$.
Then we obtain the effective Hamiltonian
\begin{eqnarray}
\label{H}
H_I&=&H_1+H_2\nonumber\\
&=&i\xi_1\hat{a}_1^\dag\hat{c}^\dag-i\xi_1^*\hat{a}_1\hat{c}+i\xi_2\hat{a}_2^\dag\hat{c}-i\xi_2^*\hat{a}_2\hat{c}^\dag.
\end{eqnarray}
The ensemble excitations and the cavity modes represent a system of three coupled
harmonic oscillators with controllable coefficients. Hamiltonian $H_1$ describes simultaneous creation or annihilation of
a photon in mode 1 and atomic spin excitation, while $H_2$ describes
the exchange of excitation quanta between mode 2 and the collective
atomic spin-waves. The effective coupling strengths $\xi_1$ and
$\xi_2$ depend on the value of the Raman transition rates.

The Hamiltonian $H_I$ commutates with the
constant of motion $N=\hat{a}_2^\dag\hat{a}_2-\hat{a}_1^\dag\hat{a}_1+\hat{c}^\dag\hat{c}$. Therefore, if the system starts from the state
$\vert 0,0\rangle_c\vert\textbf{0}\rangle_a$, where $\vert n,n\rangle_c$ is the two-mode Fock state for the cavity modes, then we have
$N=0$ at any time during the evolution. To calculate
the evolved state, we proceed by factorizing the
temporal evolution operator of the system. To this aim, we introduce the following operators $J_1=\hat{a}_1\hat{a}_1^\dag+\hat{c}^\dag\hat{c},J_2=\hat{c}^\dag\hat{c}-\hat{a}_2^\dag\hat{a}_2,J=\hat{a}_2\hat{c}^\dag,K=\hat{a}_1\hat{c},M=\hat{a}_1\hat{a}_2$, which form a closed algebra \cite{26,27}. After straightforward derivation, the evolution operator
$U(t)$ can be written in the form of a Baker-Hausdorff
equation
\begin{eqnarray}
U(t)&=&e^{\alpha_1K^\dag}e^{\alpha_2M^\dag}e^{\alpha_3J^\dag}e^{\alpha_4J_1}e^{\alpha_5J_2}e^{\alpha_6J}e^{\alpha_7K}e^{\alpha_8M}.
\end{eqnarray}
After applying $U(t)$, the time evolution of the state vector will be \cite{26,27}
\begin{eqnarray}
\label{E1}
\vert\Psi(t)\rangle&=&U(t)\vert 0,0\rangle_c\vert\textbf{0}\rangle_a\nonumber\\
&=&e^{\alpha_4}e^{\alpha_1K^\dag}e^{\alpha_2M^\dag}\vert 0,0\rangle_c\vert\textbf{0}\rangle_a\nonumber\\
&=&e^{\alpha_4}\sum_{m,n=0}^\infty\alpha_1^m\alpha_2^n\sqrt{\frac{(m+n)!}{m!n!}}\vert m+n,n\rangle_c\vert \textbf{m}\rangle_a,
\end{eqnarray}
where $e^{\alpha_4}=\frac{1}{\sqrt{1+n_1}},\alpha_1=[\frac{n_3}{1+n_1}]^{1/2},\alpha_2=[\frac{n_2}{1+n_1}]^{1/2}.$ The time dependent parameters $n_1=\langle\hat{a}_1^\dag\hat{a}_1\rangle,n_2=\langle\hat{a}_2^\dag\hat{a}_2\rangle,n_3=\langle\hat{c}^\dag\hat{c}\rangle$, which can be evaluated through solving the Heisenberg evolution of the
field operators $\hat{a}_1,\hat{a}_2,\hat{c}$ \cite{26,27}, i.e.,
$n_1=n_2+n_3,
n_2=\frac{|\xi_1|^2|\xi_2|^2}{\Theta^4}[\cos\Theta t-1]^2,
n_3=\frac{|\xi_1|^2}{\Theta^2}\sin^2(\Theta t).$
Generally the state (\ref{E1}) describes tripartite entanglement among
cavity modes and collective spin excitations. However, from the expressions for $n_i (i=1,2,3)$, we see that at the instant
$T_\pi=\pi/\Theta$, $n_1=n_2=\frac{4|\xi_1|^2|\xi_2|^2}{\Theta^4}$ and $n_3=0$. Therefore, from Eq. (\ref{E1}) we
can obtain
\begin{eqnarray}
\label{E3}
\vert\Psi(T_\pi)\rangle
&=&\frac{1}{\sqrt{1+n_1}}\sum_{n=0}^\infty[\frac{n_1}{1+n_1}]^{n/2}\vert n,n\rangle_c\vert \textbf{0}\rangle_a\nonumber\\
&=&\left(\frac{1-r^2}{1+r^2}\right)\sum_{n=0}^\infty[\frac{2r}{1+r^2}]^n\vert
n,n\rangle_c\vert \textbf{0}\rangle_a,
\end{eqnarray}
with $r=|\xi_2/\xi_1|$. This result shows that if at $t=0$ the state of the system is
$\vert\Psi(0)\rangle=\vert 0,0\rangle_c\vert \textbf{0}\rangle_a$, then at $t=T_\pi$,
the collective atomic spin excitations are decoupled from two cavity modes. Moreover,
the two cavity modes are in the state
\begin{equation}
\label{EPR}
\vert\phi\rangle_c=\left(\frac{1-r^2}{1+r^2}\right)\sum_{n=0}^\infty[\frac{2r}{1+r^2}]^n\vert
n,n\rangle_c.
\end{equation}
Obviously, state (\ref{EPR}) is a two-mode squeezed state
of the two cavity modes \cite{28}. The squeezing parameter is
$\epsilon=\tanh^{-1}(\frac{2r}{1+r^2})$, and is determined by
$|\xi_2/\xi_1|$, thus by the ratio $|\beta_2/\beta_1|$, which means the degree of squeezing and the preparation time can be
directly controlled through tuning the external classical fields. It is straightforward to generate the single mode squeezed state, if we consider the two cavity modes are identical.

To further quantify the squeezing property of the two cavity modes, we employ the two-mode relative number
squeezing parameter \cite{29} $\zeta_{12}=\sigma^2(\hat{a}_1^\dag\hat{a}_1-\hat{a}_2^\dag\hat{a}_2)/(\langle\hat{a}_1^\dag\hat{a}_1\rangle+\langle\hat{a}_2^\dag\hat{a}_2\rangle)$, where $\sigma^2(X)=\langle X^2\rangle-\langle X\rangle^2$.
$\zeta_{12}$ taking the value of 0 signifies two-mode squeezing of the cavity fields, while $\zeta_{12}=1$ means the two cavity modes are in independent states. In Figure 2(a), we display the numerical results for the time evolution of the two-mode relative number
squeezing parameter for several values of the parameter $r$. We see that ideal squeezed states between the two cavity modes occur at the instant $T_\pi=\pi/\Theta$. When the squeezed
state is generated at the time of $T_\pi$, we switch off the lasers and decouple the atoms to the cavity. Then the squeezed state can be preserved
until the cavity fields are coupled out.

The squeezing of the output fields can be measured by the standard homodyne detection. We consider the squeezing
properties of the outgoing cavity fields.
Define $I_+^{\mbox{out}}=\frac{1}{\sqrt{2}}(\hat{a}_1+\hat{a}_1^\dag-\hat{a}_2-\hat{a}_2^\dag)$, and $I_-^{\mbox{out}}=-\frac{i}{\sqrt{2}}(\hat{a}_1-\hat{a}_1^\dag+\hat{a}_2-\hat{a}_2^\dag)$, corresponding to the
difference between the amplitude
quadratures, and the sum of the phase quadratures of
the two cavity modes respectively. Then the squeezing spectrum for the output modes
can be defined as \cite{10}
\begin{equation}
\langle I_{\pm}^{\mbox{out}}(\omega)I_{\pm}^{\mbox{out}}(\omega')+I_{\pm}^{\mbox{out}}(\omega')I_{\pm}^{\mbox{out}}(\omega)\rangle
=2S_{\pm}(\omega)\delta(\omega+\omega'),
\end{equation}
with $I(\omega)$ being Fourier transformation of $I$.
The squeezing spectrum can take value from 0 to 1. The shot noise level corresponds to $S_{\pm}(\omega)=1$, while the two-mode squeezing
corresponds to $S_{\pm}(\omega)<1$.
Employing the input-output theory and quantum Langevin equations \cite{1}, we can evaluate the squeezing spectrum $S(\omega)$. Figure 2(b) shows the numerical calculations of the squeezing spectrum under several
values of the parameter $\Theta/\kappa$. We see that the properties of the spectrum
are mainly related to the ratio $\Theta/\kappa$. In the regime of $\Theta>\kappa$, one can find three minima in the squeezing spectrum at $\omega=0,\pm\Theta$, which signify three separated regions of narrow-band squeezing. In the regime of $\Theta\sim\kappa$, however, the three minima
merge into a single broad one, centered around $\omega=0$. Moreover, in this case one can find nearly perfect squeezing in the center. When $\Theta<\kappa$, the spectrum displays one very narrow bandwidth around $\omega=0$, and two-mode squeezing is significantly worsened.
\begin{figure}[htb]
\centerline{\includegraphics[bb=47 47 248 149,totalheight=1.4in,clip]{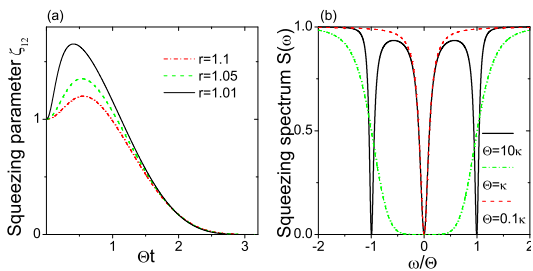}}
\caption{(Color online) (a) The parameter $\zeta_{12}$ vs the time under $r=1.1,1.05,1.01$. (b) Squeezing spectrum $S(\omega)$ when $\Theta=10\kappa,\kappa,0.1\kappa$ (in units of $\Theta$).}
\end{figure}

In the discussions above, we have assumed that a cloud of atoms can be trapped close to a transmission line resonator and prepared in the ground state, and the atom-field coupling strengths are uniform
through the atomic cloud. We now analyze these assumptions are reasonable with the state-of-the-art technology in experiment. With the help of atom-chip technology, an ensemble of cold atoms and a superconducting transmission line resonator can be integrated in a hybrid device on a single superconducting atomchip \cite{17,18,19,20}. The preparation of the initial atomic state can be accomplished through the well-developed optical pumping and adiabatic population transfer techniques. We assume modest atomic densities, such that atomic interactions can safely be neglected when they are in the ground state. In such a case, mechanical interaction between atoms can be avoided, and the internal degree of freedom of atoms can be decoupled from atomic motion. A typical ensemble of cold Rb atoms confined in an elongated trap on the atomchip possesses a transverse extension of about 1 $\mu$m and a length of up to several millimeters. Therefore, the variation of the microwave field of the resonator is neglectable on these length scales. If we fix the atomic cloud parallel to the transmission line resonator, we can neglect the change in the microwave field over the atomic cloud and assume uniform coupling constants for all the atoms.

For experimental implementation of this protocol, a promising
candidate for the atoms is $^{87}$Rb coupled to a stripline
resonator \cite{19}. The dominant interactions with a microwave field for Rb atoms cooled in the ground state are the magnetic
dipole transitions between the atomic hyperfine states of the $5S_{1/2}$ground state \cite{30}. We choose $\vert F=1,m_F=-1\rangle=\vert g\rangle$, $\vert F=1,m_F=0\rangle=\vert h\rangle$, and the intermediate states $\vert F=2,m_F=0\rangle=\vert e_1\rangle$ and $\vert F=2,m_F=-1\rangle=\vert e_2\rangle$ \cite{30}. If we replace the intermediate states $\vert e_1\rangle$ and $\vert e_2\rangle$ by a single level $\vert e\rangle$, then we can choose the atomic levels such that $\vert F=1,m_F=-1\rangle=\vert g\rangle$, $\vert F=1,m_F=1\rangle=\vert h\rangle$, and $\vert F=2,m_F=0\rangle=\vert e\rangle$. Using the experimental setup demonstrated in Ref. \cite{29} and other technology of circuit cavity QED \cite{15}, an ensemble of cold Rb atoms can be
positioned near the surface of a stripline
resonator. For an atomic ensemble of $N\sim10^6-10^8$ $^{87}$Rb atoms trapped several $\mu$m above the surface of a stripline cavity, a collective coupling strength of $\sqrt{N}g/2\pi\sim40-400$
kHZ can be obtained \cite{19}, which dominates the cavity decay $\kappa/2\pi\sim7$ kHZ.
We choose the Rabi frequencies of the classic fields as $\Omega_1\sim g$, $\Omega_2\sim1.1\Omega_1$, and $\Delta\sim10g$.
For the hyperfine transition between $\vert F=1,m_F\rangle$ and  $\vert F=2,m_F\rangle$ states at a frequency of
$\nu_0=\omega_0/2\pi=6.83$ GHz, the frequencies of the classical fields and the cavity modes can be $\omega_1\sim\omega_2\sim\nu_0-\Delta$, $\nu_1\sim\nu_0-\Delta,\nu_2\sim2\nu_1$.
With the chosen parameters, we can obtain the angular frequency
$\Theta/2\pi\sim10$ kHZ, and the time to prepare the squeezed state $T_{\pi}\sim50\mu$s, with the squeezing degree $\epsilon\sim3$ ($r\sim1.1$) and average number of $110$ photons per mode.

At this stage, let us discuss the effect of thermal photons and estimate the reasonable parameter range in which the contribution of thermal photons can be neglected. To this end, we adopt the discussions in Ref. \cite{19}, where they consider the elimination of thermal cavity photons in an internal very cold Rb gas residing in one hyperfine state. This can be applied to our setup safely, because in the squeezing scheme the atoms also keep in the lower hyperfine state when coupling to the cavity modes. Moreover, in Ref. \cite{19} they also discuss the coupling of cold Rb atoms with several internal states to the stripline cavity, which in essence is the coupling scheme employed here. The number of thermal photons can be estimated as $\bar{n}_T=[\exp(\frac{\hbar\omega_0}{k_BT})-1]^{-1}$. To maintain an empty cavity thus requires $\bar{n}_T\ll1$. With $\omega_0/2\pi=6.83$ GHz corresponding to a temperature $T\sim350$ mK, cooling to
below 100 mK is thus required to eliminate the thermal photons. To this aim, one can employ a perfectly polarized BEC with all atoms in
the lower hyperfine state coupling to the stripline cavity. For $^{87}$Rb an effective internal temperature as low as $30$ mK can be obtained. Coupling these two systems will allow an energy flow
towards the ensemble of ultracold atoms. The
photon absorption rate from the cavity into the atomic
ensemble can be estimated as $\gamma_c/2\pi\sim g^2N/\gamma_a2\pi$ \cite{19}, with $\gamma_a^{-1}$ the lifetime
of the untrapped upper state. With a heating rate $R_h\sim\kappa\bar{n}_T$, the suppression
of the thermal photons is then given by $\kappa/(\gamma_c+\kappa)$. So we can remove thermal photons from the mode as long as $\gamma_c\gg\kappa$, which can be fulfilled for the parameters given above.

We now briefly discuss how this scheme can be used as the building blocks for quantum network with continuous variables. We consider two distant clouds of cold atoms A and B coupled to a transmission line cavity. The level structure of the atoms is shown in Figure 1(b). After the cavity modes are prepared in the two-mode squeezed state, we hope to use this squeezing source to prepare the two atomic ensembles in a squeezed state. To accomplish this task, we assume the following coupling schemes for ensembles A and B, $\vert g\rangle\leftrightarrow\vert e_1\rangle\leftrightarrow\vert h\rangle$ for ensemble A and $\vert g\rangle\leftrightarrow\vert e_2\rangle\leftrightarrow\vert h\rangle$ for ensemble B. Then using the stimulated
Raman adiabatic passages \cite{31}, we can accomplish the following process, $\left(\frac{1-r^2}{1+r^2}\right)\sum_{n=0}[\frac{2r}{1+r^2}]^n\vert
n,n\rangle_c\vert \textbf{0}\rangle_A\vert \textbf{0}\rangle_B\rightarrow\left(\frac{1-r^2}{1+r^2}\right)\sum_{n=0}[\frac{2r}{1+r^2}]^n\vert
0,0\rangle_c\vert \textbf{n}\rangle_A\vert \textbf{n}\rangle_B$, where $\vert \textbf{n}\rangle_{A/B}=1/\sqrt{n!}c^{\dag n}_{A/B}\vert \textbf{0}\rangle_{A/B}$, $\vert\textbf{0}\rangle_A=\vert h_1h_2...h_N\rangle_A,\vert\textbf{0}\rangle_B=\vert g_1g_2...g_N\rangle_B$ $c^\dag_{A}=1/\sqrt{N_A}\sum_{j=1}^N\vert g_j\rangle\langle h_j\vert,c^\dag_{B}=1/\sqrt{N_B}\sum_{j=1}^N\vert h_j\rangle\langle g_j\vert$. Therefore, we are able to prepare the atomic ensembles in a squeezed state using the produced squeezed fields.  Entangled distant atomic ensembles are the building blocks for quantum network \cite{32}. Thus this protocol opens up the possibility for implementing on-chip quantum network.

It is worth emphasizing that, comparing to the existing proposals, the present proposal possesses the following distinct features : (i) This
scheme is immune
to charge noise and could constitute qubits with much
longer coherence times, due to utilizing the strong magnetic coupling of atomic spins to cavity modes \cite{19}. (ii) It makes use of ultracold atoms coupling to transmission line resonators. Therefore, the well-developed techniques for detecting and manipulating the ground electronic
(hyperfine) states of cold atoms can be employed here. (iii) These solid-state circuits allow dense integration and scaling. If integrating both systems on a single atomchip, this will
offer a promising platform for the implementation of on-chip
quantum information processing.

In conclusion, we have proposed an efficient method for controllably generating field squeezing with cold atoms coupling to
a superconducting transmission line cavity. We show that, with the
strong magnetic coupling between the collective atomic spins and the cavity modes, single mode or two mode squeezed states can be generated through coherent control on the dynamics of the system. Our proposal may allow to perform on-chip quantum information processing with continuous variables
in the microwave regime.

This work is supported by the National Nature Science Foundation of China under Grant Nos.60778021 and the National
Key Project of Basic Research Development under Grant No.2010CB923102. P.-B.L. acknowledges the support from the New Staff Research Support Plan of Xi'an Jiaotong University under No.08141015 and
the quite useful discussions with Hong-Yan Li.




\end{document}